\documentclass[prl,twocolumn,showpacs,preprintnumbers,amsmath,amssymb,superscriptaddress,nofootinbib]{revtex4-1} 
\usepackage{mathrsfs, epsfig, graphicx, url, dcolumn, bm}
\usepackage{hyperref}
\usepackage{hypernat}

\newcommand{\newsec}[1]{\vspace{0.25cm}\noindent{\bf \emph{#1}}.\hspace{0.25cm}}
\newcommand{\newsubsec}[1]{\vspace{0.15cm}\noindent $\bullet$ \emph{#1}.\hspace{0.25cm}}

\begin{document}

\title{Will Nonlinear Peculiar Velocity and Inhomogeneous Reionization Spoil 21cm Cosmology from the Epoch of Reionization?}

\author{Paul R. Shapiro}
\email{shapiro@astro.as.utexas.edu (PRS)}
\affiliation{Department of Astronomy and Texas Cosmology Center, University of Texas, Austin, TX 78712, USA}
\author{Yi Mao}
\email{mao@iap.fr (YM)}
\affiliation{Department of Astronomy and Texas Cosmology Center, University of Texas, Austin, TX 78712, USA}
\affiliation{UPMC Univ Paris 06, CNRS, Institut Lagrange de Paris, Institut d'Astrophysique de Paris, UMR7095, 98 bis, boulevard Arago, F-75014, Paris, France}
\author{Ilian T. Iliev}
\affiliation{Astronomy Centre, Department of Physics \& Astronomy, Pevensey II Building, University of Sussex, Falmer, Brighton BN1 9QH, UK} 
\author{Garrelt Mellema}
\affiliation{Department of Astronomy \& Oskar Klein Centre, AlbaNova, Stockholm University, SE-106 91 Stockholm, Sweden }
\author{Kanan K. Datta}
\affiliation{Department of Astronomy \& Oskar Klein Centre, AlbaNova, Stockholm University, SE-106 91 Stockholm, Sweden }
\author{Kyungjin Ahn}
\affiliation{Department of Earth Sciences, Chosun University, Gwangju 501-759, Korea}
\author{Jun Koda}
\affiliation{Centre for Astrophysics \& Supercomputing, Swinburne University of Technology, Hawthorn, Victoria 3122, Australia}

\date{Submitted to Phys.~Rev.~Lett.\ November 12, 2012; Accepted February 28, 2013} 




\begin{abstract}

The 21cm background from the epoch of reionization is a promising cosmological probe: line-of-sight velocity fluctuations distort redshift, so brightness fluctuations in Fourier space depend upon angle, which linear theory shows can separate cosmological from astrophysical information. Nonlinear fluctuations in ionization, density and velocity change this, however. The validity and accuracy of the separation scheme are tested here for the first time, by detailed reionization simulations. The scheme works reasonably well early in reionization ($\lesssim 40\%$ ionized), but not late ($\gtrsim 80\%$ ionized). 

\end{abstract}


\setcounter{footnote}{0}

\bigskip

\maketitle

\newsec{Introduction} 
Neutral hydrogen atoms in the intergalactic medium (IGM) at high redshift contribute a diffuse background of redshifted 21cm-line radiation which encodes a wealth of information about physical conditions in the early universe at $z>6$, during and before the epoch of reionization (EOR). 
To derive {\it cosmological} information from this, however, we must be able to {\it separate} the dependence of the signal on the background cosmology (i.e.\ total matter density fluctuations) from that on the complex {\it astrophysical} processes that cause the thermal and ionization state of the intergalactic gas to fluctuate. 
The anisotropy introduced by the peculiar velocity of the gas, induced by structure formation, is the key to this separation. 

According to linear perturbation theory, the three-dimensional power spectrum of the 21cm brightness temperature fluctuations (hereafter ``21cm power spectrum'') can be expressed as a sum of terms which depend on different powers of the cosine, $\mu_{\bf k}$, of the angle between the line-of-sight (LOS) ${\bf n}$ and the wavevector of a given Fourier mode ${\bf k}$ \cite{Barkana05}. Different terms represent contributions from different sources of fluctuations, including fluctuations in the total matter density, velocity, and hydrogen ionized fraction, and thereby in principle provide a means of separating the effects of {\it cosmology} and {\it astrophysics}. In particular, future measurements, it is proposed \cite{Barkana05}, can be used to fit this theoretical dependence of the power spectrum on $\mu_{\bf k}$ to extract the total matter density power spectrum -- the {\it cosmological} jewel. 

The success of this approach, however, depends upon the validity of the linear $\mu_{\bf k}$-decomposition. While fluctuations in the total matter density at high redshift are of linear amplitude on large scales, the nonlinearity of small-scale structure in density, velocity and reionization patchiness can leave its imprint on the 21cm signal, which might result in nonlinear distortion in the 3D 21cm power spectrum and so spoil the linear $\mu_{\bf k}$-decomposition. In what follows, we will examine this question. We will assess the accuracy of this method for deriving cosmological information from the 21cm background by using the results of a new large-scale N-body+radiative transfer simulation of cosmic reionization as mock data. 
Our simulation volume $(425\,{\rm Mpc}/h)^3$ (comparable to LOFAR\cite{lofar} survey volume) is large enough to make the sampling errors smaller than the systematic errors.

\newsec{The 21cm power spectrum anisotropy in redshift-space} 
The observed frequency 
reflects both the cosmological redshift $z_{\rm cos}$ from the time of emission and the Doppler shift associated with the peculiar radial velocity $v_\parallel$ there. 
In the ``distorted'' comoving coordinate system known as {\it redshift} space, the position of the emitter is the {\it apparent} comoving position if the redshift is interpreted as cosmological only, which shifts the {\it real} comoving coordinate ${\bf r}$ along the LOS to ${\bf s} = {\bf r} + \frac{1+z_{\rm cos}}{H(z_{\rm cos})}\,v_\parallel \, {\bf n}$ (where $H(z)$ is the Hubble parameter). 
Henceforth, superscripts ``r'' and ``s'' denote quantities in real- and redshift-space, respectively, and we will write $z$ for $z_{\rm cos}$. 

The 21cm signal in redshift space can be modeled as follows, based on different assumptions on peculiar velocity and reionization patchiness. 

\newsubsec{Linear scheme (linear velocity-density relation, linearized neutral fraction fluctuations)}
The linear scheme was originally proposed in the context of linear perturbation theory  \cite{Barkana05} and later re-derived with weaker assumptions \cite{Mao12} which are: (1) the velocity (${\bf v}^r$) and total density fluctuations ($\delta^r_{\rho}$) satisfy the linear relation, $\widetilde{v^r_\parallel}({\bf k}) = i\left(\frac{H(z)}{1+z}\right)\widetilde{\delta^r_{\rho}}({\bf k}) \mu_{\bf k}/k$, (2) the baryon distribution traces the CDM, (3) the peculiar velocity, the hydrogen density fluctuation ($\delta^r_{\rho_{\rm H}}$), and the {\it neutral fraction} fluctuation ($\delta^r_{x_{\rm HI}}$) are all linearized. Under these assumptions, the 3D 21cm power spectrum can be expanded in polynomials of $\mu_{\bf k} \equiv {\bf k} \cdot {\bf n}/|{\bf k}|$, 
\begin{equation}
P_{\Delta T}^{s,{\rm lin}} ({\bf k},z) = P_{\mu^0}(k,z) + P_{\mu^2}(k,z) \mu_{\bf k}^2 + P_{\mu^4}(k,z) \mu_{\bf k}^4 \,.
\label{eqn:lin-scheme}
\end{equation}
While the 0$^{\rm th}$- and 2$^{\rm nd}$-moments are ``contaminated'' by power spectra due to reionization and/or spin temperature, 
the 4$^{\rm th}$-moment depends only on $P^{r,{\rm total}}_{\delta_{\rho},\delta_{\rho}}$ (total density power spectrum) and $\bar{x}_{\rm HI,m}(z)$ (mass-weighted global neutral fraction), 
\begin{equation}
P_{\mu^4}(k,z) = \widehat{\delta T}_b^2(z) P^{r,{\rm total}}_{\delta_{\rho},\delta_{\rho}}(k,z)\,.\label{eqn:BL05-mu4}
\end{equation} 
Here $\widehat{\delta T}_b$ is the mean 21cm signal,   
$\widehat{\delta T}_b (z) = (23.88\,{\rm mK}) \left(\frac{\Omega_{\rm b}h^2}{0.02}\right)\sqrt{\frac{0.15}{\Omega_{\rm M} h^2}\frac{1+z}{10}}\,\bar{x}_{\rm HI,m}(z)$
\footnote{Here we focus on the limit where the spin temperature $T_s^r \gg T_{\rm CMB}$, valid soon after reionization begins. As such, we can neglect the dependence on spin temperature, but our discussion can be readily generalized to the case of finite $T_s^r$.}. 
In principle, then, cosmological information can be extracted from the 21cm signal by fitting the measured $P_{\Delta T}^{s} ({\bf k},z)$ to equation (\ref{eqn:lin-scheme}) to isolate the 4$^{\rm th}$ moment.

\newsubsec{Quasi-linear $\mu_{\bf k}$-decomposition scheme (linear velocity-density relation, linearized neutral overdensity)}
The assumption of $\delta^r_{x_{\rm HI}} \ll 1 $ breaks down when $\bar{x}_{\rm HI,m} < 0.5$
\cite{Lidz07}. 
Fortunately, equations~(\ref{eqn:lin-scheme}) - (\ref{eqn:BL05-mu4}) can still hold, 
if we adopt the same assumptions (1) and (2) as in the linear scheme, but assume peculiar velocity and the {\it neutral density} fluctuation $\delta^r_{\rho_{\rm HI}} = \delta^r_{\rho_{\rm H}} + \delta^r_{x_{\rm HI}} + \delta^r_{\rho_{\rm H}} \, \delta^r_{x_{\rm HI}}$, as opposed to $\delta^r_{x_{\rm HI}}$ alone, are linearized. In this so-called quasi-linear $\mu_{\bf k}$-decomposition formalism \cite{Mao12}, only lower moments 
differ from the linear scheme prediction. 

\newsubsec{Fully-nonlinear scheme (nonlinear velocity, nonlinear neutral overdensity)}
In the optically-thin approximation, two nonlinear effects of peculiar velocity must be taken into account: 
(1) the 21cm brightness temperature is corrected for velocity gradient \cite{Barkana05}, 
$\delta T_b^s ({\bf s}) = \delta T_b^r ({\bf r}) = \widehat{\delta T}_b (z) \, \frac{1+\delta^r_{\rho_{\rm HI}}({\bf r})}{\left|1+\delta^r_{\partial_r v}({\bf r})\right|}$,    
where $\delta^r_{\partial_r v}({\bf r}) \equiv \frac{1+z}{H(z)} \frac{dv_\parallel}{dr_\parallel }({\bf r})$ is the gradient of proper radial peculiar velocity along the LOS, normalized by $\frac{H}{1+z}$;
(2) when the {\it real}-space comoving coordinates ${\bf r}$ are mapped to {\it redshift}-space coordinates, volume elements are also resized according to $\delta V^s({\bf s}) = \delta V^r({\bf r}) \left|1+\delta^r_{\partial_r v}({\bf r})\right|$. Fortunately, the combined effect allows us to compute the 21cm brightness temperature in redshift-space with a simple formula \cite{Mao12}, 
$\delta T_b^s ({\bf s}) 
= \widehat{\delta T}_b (z) \, \left[1+\delta^s_{\rho_{\rm HI}}({\bf s})\right]$, 
where $\delta^s_{\rho_{\rm HI}}({\bf s}) = n^s_{\rm HI}({\bf s})/\bar{n}_{\rm HI}(z)-1$ is the neutral overdensity in redshift-space. 

\begin{figure*}[ht]
\includegraphics[width=0.9\textwidth]{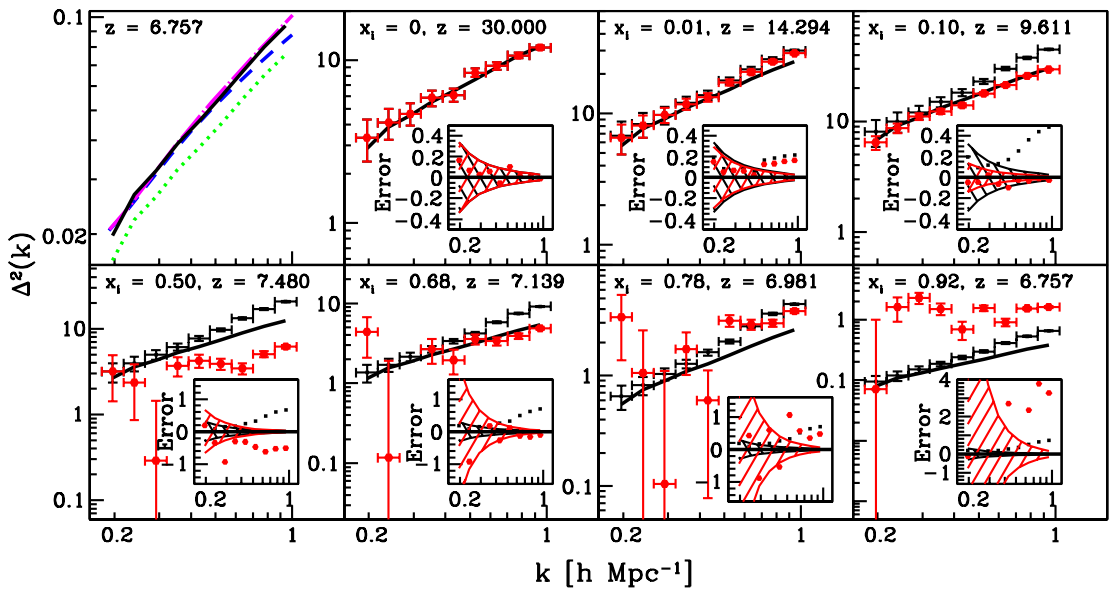}
\caption{
Top left panel: the matter density power spectrum, $\Delta^2_{\delta\delta}= k^3 P_{\delta\delta}(k)/2\pi^2$ (unitless), from N-body results for {\it total} matter (solid/black) and for {\it IGM} matter (green/dot), linear total density (dashed/blue), and 3PT total density (dot-dashed/magenta), at $z=6.757$. 
Other panels: the total matter density power spectrum (multiplied by $\widehat{\delta T}_b^2(z)$) as ``measured'' by fitting the mock data (i.e.~reionization simulation) for the 3D 21cm power spectrum to the $\mu_{\bf k}$-decomposition, $\Delta^2_{\mu^4}= k^3 P_{\mu^4}(k)/2\pi^2$ (in ${\rm mK}^2$), at different phases of reionization ($\bar{x}_{i,m} = 0$, 0.01, 0.10, 0.50, 0.68, 0.78, and 0.92, respectively). Red data points are reionization simulation results. Black data points assume homogeneously-ionized IGM with same $\bar{x}_{i,m}$ as reionization simulation. Solid black curves are the expectation from linear scheme, $P_{\mu^4}^{\rm linear}$, 
i.e.~equation~(\ref{eqn:BL05-mu4}) evaluated using {\it total} density power spectra from N-body simulation (multiplied by $\widehat{\delta T}_b^2(z)$). 
Plotted in inset are {\it fractional systematic errors}, i.e.\ $ (P_{\mu^4}^{\rm best-fit}-P_{\mu^4}^{\rm linear})/P_{\mu^4}^{\rm linear}$. 
We also plot the sampling errors (due to the simulation volume) for the best-fit 4$^{\rm th}$-moment, $\sigma(P_{\mu^4})$, as error bars, and the {\it fractional} sampling errors, $\sigma(P_{\mu^4})/P_{\mu^4}^{\rm linear}$, as shaded regions in inset. 
}
\label{fig:result}
\end{figure*}

\newsec{Angular separation of 3D power spectrum} 
The moments of 3D power spectrum 
can be decomposed using the $\chi^2$-fit as follows. For a given LOS, 3D modes $P_{\Delta T}^s({\bf k})$ with the same $\mu_{\bf k}$ and same $k$ but different azimuthal angle are averaged to give a measure of $\overline{P_{\Delta T}^s}(k,\mu_{\bf k})$ and the associated sampling variance $\sigma_{P}^2(k,\mu_{\bf k}) = \left(2/N_{\mu_{\bf k},k}\right) \,\overline{P_{\Delta T}^s}^2(k,\mu_{\bf k})$, where $N_{\mu_{\bf k},k}$ is the number of modes with the same $\mu_{\bf k}$ and $k$. 
For the 3D grid from the simulated data, 
we further combine the measures along three different LOS, tripling the number of modes. Next, modes are grouped in spherical $k$-shells with the width $\Delta k/k = 0.186$ (chosen as a trade-off between minimal mean $\chi^2$ and minimal numerical noise, but our results below are insensitive to this value). 
For each shell (the $k$-dependence is implicit below), we shall fit the data set $\left\{ \mu_{i},\overline{P_{\Delta T}^s}(\mu_{i}), \sigma_{P}(\mu_{i}) \right\}$ (where $i$ runs through all modes within the shell) with the ansatz $P_{\Delta T}^s(\mu) = \sum_{j=1}^{3} a_j X_j(\mu)$, where basis functions $X_j(\mu) = \{1, \mu^2, \mu^4\}$ and coefficients $a_j = \{P_{\mu^0},P_{\mu^2},P_{\mu^4}\}$, for $j=1,2,3$, respectively. 
We employ the standard General Linear Least Squares method (see, e.g. \cite{numrec}) for the $\chi^2$-fit. 
This results in best-fit coefficients $a_j = \sum_{k=1}^3 C_{jk}\,\beta_k$ with associated error estimates $\sigma(a_j) = \sqrt{C_{jj}}$. Here we define the $3\times 3$ matrix $\alpha_{jk} = \sum_{i} \, X_j(\mu_i) X_k (\mu_i)/\sigma_{P}^2(\mu_{i}) $, whose inverse is the covariance matrix $C=\alpha^{-1}$, and a 3-vector $\beta_j = \sum_{i}\,\overline{P_{\Delta T}^s}(\mu_{i}) X_j(\mu_i)/\sigma_{P}^2(\mu_{i}) $. 

\newsec{Mock data from a reionization simulation}
We perform a new large-scale, high-resolution N-body simulation of the $\Lambda$CDM universe (performed with the {\small CubeP$^3$M} code \cite{Iliev08,Harnois-Deraps12}) in a comoving volume of 
$425$\,Mpc/$h$ on each side using $5488^3$ (165 billion) particles. To find halos, we use the spherical overdensity method and require them to consist of at least 20 N-body particles; this implies a minimum halo mass of $10^9\,M_\odot$. We use subgrid modeling to compute the halo population with mass between $10^8 - 10^9\,M_\odot$. 
Assuming that the gas traces the CDM particles exactly, we grid the density and velocity in the IGM (i.e.~halo mass excluded) on a $504^3$ grid using SPH-like smoothing with an adaptive kernel. Halo lists and density fields are then processed with the radiative transfer code {\small C$^2$Ray} \cite{Mellema06}. Each halo releases $f_\gamma$ ionizing photons per baryon per $\Delta t = 11.5$\,Myrs, with $f_\gamma=10$ ($f_\gamma=2$) for halos below $10^9\,M_\odot$ (above $10^9\,M_\odot$), respectively. To incorporate feedback from reionization, halos less massive than $10^9\,M_\odot$ located in ionized regions do not produce any photons. We refer the readers to \cite{Iliev12a} 
and Iliev et al.~({\it in prep.}) 
for more details of this simulation. 
The background cosmology is consistent with the {\it WMAP} seven-year results \cite{Komatsu11}: 
$\Omega_\Lambda=0.73, \Omega_{\rm M}=0.27, \Omega_{\rm b}=0.044, h=0.7, \sigma_8=0.8, n_\mathrm{s}=0.96$. 

We then calculate the nonlinear 3D 21cm power spectrum, using the {\small MM-RRM} scheme in \cite{Mao12} for mapping the nonlinear density, velocity and ionization grid data from the simulation in real space onto a redshift-space grid, and separate out the best-fit 4$^{\rm th}$-moment using the aforementioned angular separation prescription. 
The linear and quasi-linear schemes both predict that this 4$^{\rm th}$-moment
should be given by equation~(\ref{eqn:BL05-mu4}), which we test by evaluating the r.h.s.~of equation~(\ref{eqn:BL05-mu4}) directly from the simulation N-body results, for comparison.
Some preliminary results were previously summarized by us in \cite{Mao10}.

\newsec{Results and discussions}
In Figure~\ref{fig:result}, we plot the best-fit 4$^{\rm th}$-moment of the fully nonlinear power spectrum, and the benchmark linear expectation. 
Note that, while the mock 21cm signal is from the IGM, as is the observed signal, the {\it total} density power spectrum is expected in the linear scheme. (There is a difference between the {\it total} and {\it IGM} density power spectra, as in Figure~\ref{fig:result}, resulting from the exclusion of particle mass in halos when computing the IGM density.) 
We focus on the range of wavenumbers,
$0.2 < k < 1\,h\,{\rm Mpc}^{-1}$, for which $k$ is large enough
to avoid a large sampling variance but small
enough to avoid a large aliasing effect. 

\newsubsec{Consistency check of the decomposition pipeline} 
We confirm that: (1) the total density power spectrum from N-body simulation agrees with the linear power spectrum (from CAMB\cite{Lewis02}) at large scales $k \lesssim 0.5\,h\,{\rm Mpc}^{-1}$. At smaller scales, it agrees with the result from third-order perturbation theory (3PT)\cite{Jeong06}, but is enhanced relative to the linear power spectrum. (2) The best-fit 4$^{\rm th}$-moment agrees with the total density power spectrum at $z=30$ when the IGM is everywhere neutral and the density fluctuations are of linear amplitude.

\newsubsec{Effect of IGM nonlinear density/velocity fluctuations} 
For diagnostic purposes, we first investigate the case in which the ionized fraction at each point in the IGM is set equal to the mass-weighted global ionization fraction $\bar{x}_{i,m}$ in the reionization simulation at that redshift.
In this case, the best-fit 4$^{\rm th}$-moment at different redshifts is enhanced with respect to the total density power spectrum, and the deviation increases from 20\% ($z\simeq 14$) to 70\% ($z\simeq 7$), as structure formation proceeds.
These results show that nonlinear density and velocity fluctuations cause the $\mu_{\bf k}$-decomposition to make a systematic error, not caused by ionization patchiness.

\newsubsec{Effect of inhomogeneous reionization and velocity fluctuations} 
Early in reionization, the best-fit 4$^{\rm th}$-moment for inhomogeneous
reionization is suppressed relative to that for the homogeneously
partially-ionized case.  This is because fluctuations in neutral
fraction and density anticorrelate in a universe reionized ``inside-out,''
i.e.~overdense regions are ionized earlier on average than underdense
regions. 
This anticorrelation affects the anisotropy of the 21cm power spectrum through the coupling of nonlinear ionization fluctuations with velocity fluctuations (which are, themselves, correlated with density fluctuations).   
Otherwise, reionization patchiness, alone, cannot introduce anisotropy in the 21cm power spectrum (e.g.~as in the quasi-linear $\mu_{\bf k}$-decomposition scheme). 
A more quantitative explanation will be formulated in detail in Mao et al.~({\it in prep.}).

\begin{figure}[t]
\includegraphics[width=0.3\textwidth]{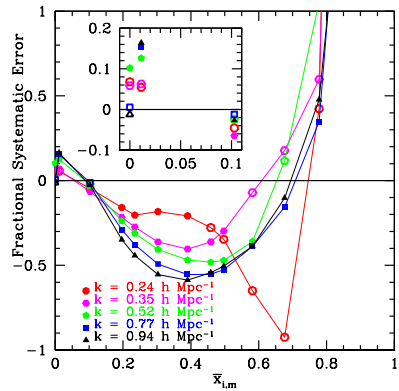}
\caption{The {\it fractional} systematic error of the separation scheme, 
in the inhomogeneous reionization case, as a function of $\bar{x}_{i,m}$, for $k = 0.24, 0.35, 0.52, 0.77, 0.94\,h\,{\rm Mpc}^{-1}$, respectively. The solid/open dots correspond to the case where the systematic error is greater/less than the sampling error of our simulation. 
The inset is zoom-in to $0\le \bar{x}_{i,m} \le 0.10$.
}
\label{fig:syserr}
\end{figure}

This effect cancels the enhancement due to nonlinear fluctuations in IGM density and velocity, alone. 
Incidentally, the fractional systematic error first crosses zero (less than 10\% at all scales) when $\bar{x}_{i,m}\approx 10\%$ ($z\simeq 9.6$ in our simulation).
Afterwards, this error grows to 60\% for $\bar{x}_{i,m} \lesssim 50\%$. As determined by the competition between these two effects, this error depends on both the reionization epoch and the scale of interest --- the smaller the ionized fraction $\bar{x}_{i,m}$ and the smaller the wavenumber $k$, the smaller the error (see Figure~\ref{fig:syserr}).

As the typical size of ionized regions grows larger than the scales plotted here, the best-fit 4$^{\rm th}$-moment at $k=0.5 - 1\,h\,{\rm Mpc}^{-1}$ becomes less suppressed after the $50\%$ ionized epoch. For {\it this} $k$-range, the net error changes sign again when $\bar{x}_{i,m} \simeq 68\%$ (with error $< 20\%$). 

At late epochs ($\bar{x}_{i,m} \gtrsim 0.8$, $z\lesssim 7$), the fractional systematic error for all scales is large, $\gtrsim 100\%$. This is due to the breakdown of the perturbative expansion, i.e.~the expansion of the 3D 21cm power spectrum in neutral density fluctuations becomes divergent when $\delta_{\rho_{\rm HI}} \gtrsim \mathcal{O}(1)$, as the ionized bubbles expand and percolate in the universe. In addition, 
while our calculation does not include the lightcone effect\cite{Datta12}, this effect becomes non-negligible at this late time, and can further squeeze the anisotropic power spectrum along the LOS, as does the redshift-space distortion. Therefore, the estimate of 100\% error here is only a lower limit to the actual error at late times. 

\newsubsec{The sampling variance} 
Our simulation volume is large enough that these systematic errors quoted above are all greater than the estimated
sampling errors (except when the systematic error crosses zero), so they represent statistically significant deviations
from the expectations of the $\mu_{\bf k}$-decomposition, rather than accidents of the particular numerical realization.
 
In the homogeneously-ionized case (denoted by the superscript ``$H$'' below), the fractional sampling error for the 4$^{\rm th}$-moment, $\sigma(P_{\mu^4})/P_{\mu^4}^{\rm linear}$, remains constant over time, which is a consequence of the unchanging number of modes. However, the fractional sampling error evolves dramatically during inhomogeneous reionization (denoted by the superscript ``$I$''), for reasons as follows. For each ring with given values of $\mu_{\bf k}$ and $k$, sampling errors of power spectra satisfy $\sigma_{P}^I/\sigma_{P}^H = \overline{P_{\Delta T}^{s,I}}/\overline{P_{\Delta T}^{s,H}} \equiv \eta(k,\mu_{\bf k})$, since $N_{\mu_{\bf k},k}$ is the same for both cases. If, for simplicity, $\eta$ is only a function of $k$, then it is straightforward to show that $\frac{\sigma^I(P_{\mu^4})/P_{\mu^4}^{\rm linear}}{\sigma^H(P_{\mu^4})/P_{\mu^4}^{\rm linear}} = \eta$. Over time $\eta$ evolves from less than to greater than unity, because: (1) in the early phase of reionization, the density power spectrum is dominant, but the $\delta^r_{\rho_{\rm H}}$-$\delta^r_{x_{\rm HI}}$ anti-correlation decreases the total power; (2) in the late phase, the ionization power dominates over the density power spectrum. 

\newsec{Conclusion} 
This {\it letter} is the first attempt to quantify in detail the
{\it intrinsic} precision of the linear scheme for extracting the cosmological matter density power spectrum from 21cm observations of the EOR. \footnote{A recent paper\cite{Majumdar12} considered the Legendre decomposition of the anisotropic 21cm power spectrum, using semi-numerical simulations, but only showed the results for monopole and quadrupole, not hexadecapole (4$^{\rm th}$-moment equivalent), due to large error bars in their simulations.}
Two effects may spoil the extraction, 
a major one due to the coupling between inhomogeneous reionization and velocity fluctuations, and a minor one due to nonlinear density and velocity fluctuations alone. 
The competition between these identifies two phases of reionization particularly interesting to cosmology --- $\bar{x}_{i,m} \simeq 68\%$, where fractional systematic error is within $20\%$ for $k = 0.5 - 1\,h\,{\rm Mpc}^{-1}$, and $\bar{x}_{i,m} \simeq 10\%$, where this error is within 10\% for {\it all} wavenumbers. 
The epoch of exact crossover is likely to depend on the reionization scenario.

We summarize our results in Figure~\ref{fig:syserr}. We see that, 
for the early phase of reionization $\bar{x}_{i,m} < 40\%$, the linear $\mu_{\bf k}$-decomposition works well for large-scale measurement $k \lesssim 0.24\,h\,{\rm Mpc}^{-1}$, with errors within $20\%$. 
At smaller scales, down to $k \simeq 0.5\,h\,{\rm Mpc}^{-1}$, errors are within $50\%$. During the intermediate phase ($\bar{x}_{i,m} \simeq 0.4 - 0.7$), using the $\mu_{\bf k}$-decomposition at the intermediate $k$-range $0.35 - 0.5\,h\,{\rm Mpc}^{-1}$ can also result in errors within $50\%$. 
However, in the late phase ($\bar{x}_{i,m} \gtrsim 0.8$), it is difficult to extract the cosmological information from 21cm observations using the $\mu_{\bf k}$-decomposition, unless possibly at very large scales $k<0.2\,h\,{\rm Mpc}^{-1}$. 


\smallskip
{\it Acknowledgements:} 
This work was supported in part by NSF grants AST-0708176 and AST-1009799, NASA grants NNX07AH09G, NNG04G177G and NNX11AE09G, Chandra grant SAO TM8-9009X, the French state funds managed by the ANR within the Investissements d'Avenir programme under reference ANR-11-IDEX-0004-02, 
the Science and Technology Facilities Council [grant numbers ST/F002858/1 and ST/I000976/1], the Southeast Physics Network (SEPNet), and the Swedish Research Council grant 2009-4088.
The authors acknowledge the Texas Advanced Computing Center
(TACC) and the National Institute for Computational Sciences (NICS) for providing HPC resources, under NSF TeraGrid grants TG-AST0900005 and TG-080028N and TACC internal allocation grant ``A-asoz''. 
Computations were performed on the GPC supercomputer at the SciNet HPC Consortium. SciNet is funded by: the Canada Foundation for Innovation under the auspices of Compute Canada; the Government of Ontario; Ontario Research Fund - Research Excellence; and the University of Toronto.

\end{document}